\documentclass{ws-procs9x6}            
\begin{document}
\title{Tetraquark mixing framework to explain two light-meson nonets}

\author{Hungchong Kim$^*$}

\address{Research Institute of Basic Science, Korea Aerospace University, \\
Goyang 412-791, Korea\\
$^*$E-mail: hungchong@kau.ac.kr\\
\vspace{3pt}
and Center for Extreme Nuclear Matters, Korea University, \\
Seoul 02841, Korea}

\begin{abstract}
In this talk, we summarize our recent works on the tetraquark mixing framework
for the two light-meson nonets in the $J^{PC} = 0^{++}$ channel,
the light nonet [$a_0(980)$, $K_0^*(700)$, $f_0(500)$, $f_0(980)$] and the heavy nonet
[$a_0(1450)$, $K_0^*(1430)$, $f_0(1370)$, $f_0(1500)$].
We briefly explain this mixing framework and present various phenomenological signatures to support this picture.
\end{abstract}


\bodymatter

\section{Tetraquark mixing framework}\label{mixing}

Recently, the tetraquark mixing framework~\cite{Kim:2016dfq,Kim:2017yur,Kim:2017yvd,Kim:2018zob,Lee:2019bwi}
has been proposed as a plausible picture to explain the two light-meson nonets 
in PDG~\cite{PDG}, the light nonet composed of $a_0(980)$, $K_0^*(700)$~\footnote{$K_0^*(700)$ was named as
$K_0^*(800)$ in the previous version of PDG~\cite{PDG_old}.
Also its average mass is now listed as 824 MeV changed from its old value of 682 MeV.},
$f_0(500)$, $f_0(980)$
constituting the lowest-lying resonances in the $J^{PC} = 0^{++}$
channel, and the heavy nonet composed of $a_0(1450)$, $K_0^*(1430)$, $f_0(1370)$, $f_0(1500)$
lying next to the lowest-lying resonances.

In constructing this framework, we reexamined the diquark-antiquark model~\cite{Jaffe}
and advocated two possible tetraquark types. 
The first tetraquark type, which is commonly adopted in tetraquark studies,
is constructed from the spin-0 diquark with the color and flavor structure of ($\bar{\bm{3}}_c, \bar{\bm{3}}_f$).
We denote this tetraquark with its spin configuration $|000\rangle$ where the first zero is the tetraquark spin,
the second the diquark spin and the third the antidiquark spin.
The second tetraquark type, which was proposed as another possibility,
is constructed from the spin-1 diquark with ($\bm{6}_c, \bar{\bm{3}}_f$). This second tetraquark
is denoted by $|011\rangle$.  The two tetraquark types differ by spin and color configurations but
both have the same flavor structure, a nonet that can be broken down to an octet and a singlet.
Common characteristics are
the quantum number, $J^{PC}=0^{++}$, possible isospins, $I=0,1/2,1$, and most importantly
the mass ordering among the octet members,
$M_{I=1} > M_{I=1/2} > M_{I=0}$.
The last one is unique in a sense that this ordering cannot be generated
from a flavor nonet constructed from a two-quark picture, $q\bar{q}$.

Physical candidates corresponding to the two tetraquark types must be sought from the resonances with $J^{PC}=0^{++}$.
Indeed, the two nonets in PDG above, the light and heavy nonets, are the possible candidates because both
satisfy the tetraquark characteristics.
Specifically, each nonet is composed of the isospin members of $I=0,1/2,1$ and
the tetraquark mass ordering is satisfied quite well for the light nonet and, for the heavy nonet,
it is still satisfied though marginally.

Then, we have the two tetraquark types $|000\rangle$, $|011\rangle$ in the one hand and
the two nonets in PDG with the tetraquark characteristics in the other hand.
Thus, it is quite tempting to match the two tetraquark types with the two nonets in PDG even though
there is a huge mass gap between the two nonets, around 500 MeV or more.
To establish a matching, a crucial observation to make is
that the two types, $|000\rangle, |011\rangle$, in each isospin channel
mix through the color-spin interaction,
$V_{CS} = v_0 \sum_{i < j}  \frac{\lambda_i \cdot \lambda_j J_i\cdot J_j}{m_i^{} m_j^{}}$,
and the expectation value, $\langle V_{CS} \rangle$, namely the hyperfine mass, forms a $2\times 2$ matrix.
The upshot is that {\it the physical resonances, namely the two nonets in PDG, can be identified by the eigenstates that
diagonalize this matrix}.
In other words, the two nonets in each isospin channel can be written as a linear combination
of $|000\rangle, |011\rangle$ and we express them collectively as
\begin{eqnarray}
|\text{Heavy~nonet} \rangle &=& -\alpha | 000 \rangle + \beta |011 \rangle \label{heavy}\ ,\\
|\text{Light~nonet} \rangle~ &=&\beta | 000 \rangle + \alpha |011 \rangle \label{light}\ ,
\end{eqnarray}
where the mixing parameters $\alpha, \beta$ are fixed by the diagonalization.
This is the tetraquark mixing framework for the two nonets in PDG.

\section{Signatures to support the tetraquark mixing framework}\label{signatures}

There are various signatures to support the tetraquark mixing framework. To explain them,
we present in Table~\ref{parameters} the hyperfine mass, $\langle V_{CS} \rangle$,
calculated in the physical basis and the mixing parameters,
$\alpha, \beta$. For the isoscalar resonances $f_0(500), f_0(980)$, we include
the flavor mixing between $\bm{8}_f$ and $\bm{1}_f$ in three different ways~\cite{Kim:2017yvd},
SSC [SU(3) symmetric case], IMC [ideal mixing case], RCF [realistic case with fitting].
Here in the table, we present the RCF results only.

\begin{table}[t]
\centering
\tbl{Hyperfine masses and the mixing parameters for the two nonets.}
{\begin{tabular}{c|c|c||c|c|c||c|c}\hline\hline
  LN & $M_{exp}$ & $\langle V_{CS} \rangle$ & HN &$M_{exp}$ & $\langle V_{CS} \rangle$  & $\alpha$ & $\beta$ \\
\hline
 $a_0(980)$   & 980  & $-488.5$ &$a_0(1450)$   & 1474 & $-16.8$ & 0.8167     & 0.5770    \\
 $K_0^*(700)$ & 824  & $-592.7$ &$K_0^*(1430)$ & 1425 & $-26.9$ & 0.8130     & 0.5822    \\
 $f_0(500)$   & 475 & $-667.5$ &$f_0(1370)$    & 1350 & $-29.2$ & 0.8136     & 0.5814    \\
 \hline
 $f_0(980)$  & 990 & $-535.1$ &$f_0(1500)$    & 1506 & $-20.1$ & 0.8157     & 0.5784    \\
\hline\hline
\end{tabular}}
\begin{tabnote}
The middle value is taken for $M_{exp}$ known with some range.\\
\end{tabnote}
\label{parameters}
\end{table}

First, the mixing parameters $\alpha, \beta$ support
our original identification of each nonet in PDG as a flavor nonet.
$\alpha, \beta$ are determined in each isospin channel separately and in principle
they may depend on isospin. But as shown in Table~\ref{parameters}, their values are almost independent of
isospin. This means, the right-hand side of Eq.~(\ref{heavy}) or Eq.~(\ref{light})
also forms a flavor nonet consistently with the corresponding nonet in the left-hand side.

Secondly, there is a strong mixing between $|000\rangle, |011\rangle$ which can explain
the huge mass gap, around 500 MeV or more, between the two nonets in PDG.
To put it more clearly, the hyperfine masses, $\langle V_{CS} \rangle$, calculated in the isovector channel
corresponding to $a_0(980),a_0(1450)$, are diagonalized to yield the eigenvalues,
\begin{eqnarray}
\begin{array}{c|lr}
 & ~|000 \rangle & |011 \rangle \\
\hline
|000 \rangle & -173.9 & -222.3\\
|011 \rangle & -222.3 & -331.5
\end{array}
\quad
&\rightarrow&
\quad
\begin{array}{l|cc}
 &~|a_0(1450) \rangle &~|a_0(980) \rangle \\
\hline
|a_0(1450) \rangle & -16.8 & 0.0\\
|a_0(980) \rangle & 0.0  & -488.5
\nonumber\
\end{array}\ .
\end{eqnarray}
The strong mixing, $\langle 000| V_{CS} |011 \rangle=-222.3$ MeV, causes a huge separation in the eigenvalues,
$-16.8-(-488.5) = 471.7$ MeV which is comparable to the experimental mass gap, $M[a_0(1450)]-M[a_0(980)] = 494$ MeV.
Indeed, in this mixing formalism, one can establish the mass splitting formula~\cite{Kim:2016dfq,Kim:2017yvd},
$\Delta M_{exp} \approx \Delta\langle V_{CS} \rangle$, between the two nonets in each isospin channel.
The experimental mass splitting is found to be consistent qualitatively with this mass splitting formula.

Thirdly, the hyperfine mass for the light nonet is negatively huge around $-500$ MeV.
As one can see from the $2\times 2$ matrix above,
this is a consequence of the mixing which pushes down substantially the hyperfine mass through the diagonalization.
This certainly helps us to understand the light nonet mass lying below
1 GeV which, without mixing, seems too low to be the tetraquark mass.
On the other hand, $\langle V_{CS} \rangle$ for the heavy nonet
is very small around $-20$ MeV so the heavy nonet mass is not far from $\sim 4 m_q$.

Fourthly, Table~\ref{parameters} shows $\alpha > \beta$ indicating that the resonances
in the light nonet, through Eq.~(\ref{light}), have more probability to stay
in $|011\rangle$ rather than in $|000\rangle$.  This is quite surprising because
the light nonet member is often believed to have the $|000\rangle$ configuration only.
The result, $\alpha > \beta$, in fact, originates from the fact that the second tetraquark $|011\rangle$ is
more compact than $|000\rangle$ when the binding is calculated from the color-spin interactions
not only for the diquark (antidiquark) but also for the other pairs composed of $q\bar{q}$.
For the isovector case, in particular, we have
$\langle 011| V_{CS} |011 \rangle=-331.5$ MeV which is more negative than
$\langle 000| V_{CS} |000 \rangle=-173.9$ MeV.  Our finding here is also supported by
a QCD sum rule calculation performed for $a_0(980)$~\cite{Lee:2019bwi} using an interpolating field involving the two tetraquark types.

Fifthly, our hyperfine masses, $\langle V_{CS} \rangle$, can give a partial explanation for the marginal mass ordering
seen in the heavy nonet.
In Table~\ref{parameters}, one can see that $\langle V_{CS} \rangle$ has the same ordering as the mass
$M_{I=1} > M_{I=1/2} > M_{I=0}$ both for the light and heavy nonets.
This means, $\langle V_{CS} \rangle$ is also responsible for the mass ordering in addition to the quark masses.
But the hyperfine splitting is much narrower in the heavy nonet than in the light nonet.
For example, we have the hyperfine splitting in the heavy nonet,
$\langle V_{CS} \rangle[a_0(1450)] - \langle V_{CS} \rangle[K^*_0(1430)]\approx 10$ MeV
which is much smaller than the corresponding splitting in the light nonet,
$\langle V_{CS} \rangle[a_0(980)] - \langle V_{CS} \rangle[K^*_0(700)]\approx 104$ MeV.
Thus, the mass splitting is narrower in the heavy nonet by the reduction of the hyperfine mass splitting.

Another signature can be found from fall-apart decay of the tetraquarks.
In this decay, two $q\bar{q}$ in a tetraquark simply fall apart into two mesons.
This decay is allowed because the tetraquarks, $|000\rangle$ and $|011\rangle$,
have a component with two color-singlet, $\bm{1}_c\otimes\bm{1}_c$, when the wave functions
are rearranged into two $q\bar{q}$ from the diquark-antidiquark ($qq\bar{q}\bar{q}$) configuration.
There are two modes depending on the two mesons in the final state,  the PP mode for two pseudoscalars
and the VV mode for two vectors.
The coupling strengths are calculated by recombining the color and spin configurations of $|000\rangle$, $|011\rangle$,
into two $q\bar{q}$.
From this recombination, we found that the PP mode is enhanced in the light nonet but suppressed in the heavy nonet.
The relative enhancement factor in the couplings is about 4.
We tested this prediction from the
ratios~\footnote{The ratios eliminate the unknown dependence from the overall constant.} of partial widths from
$a_0(980), a_0(1450)$ in comparison with the two experimental analyses, one by Bugg~\cite{Bugg:2008ig} and the other
based on PDG data.  They agree very well~\cite{Kim:2017yur},
\begin{eqnarray}
\begin{array}{c|c|c|c}
\text{Ratio}&\text{Ours} & \text{Bugg} & \text{PDG} \\
\hline
\frac{\!\!\!\Gamma [a_0(980)\rightarrow \pi \eta]}{\Gamma [a_0(1450)\rightarrow \pi \eta]}&2.51\text{--} 2.54 & 2.53 &2.93 \text{--} 3.9  \\[2mm]
\frac{\!\!\!\Gamma [a_0(980)\rightarrow K\bar{K}]}{\Gamma [a_0(1450)\rightarrow K\bar{K}]}&0.52\text{--} 0.89 & 0.62 &0.61 \text{--} 0.81
\nonumber
\end{array}
\ .
\end{eqnarray}
For the VV mode, we found the opposite trend. It is enhanced in the heavy nonet but suppressed in the light nonet.
The relative enhancement factor in the heavy nonet is about 15~\cite{Kim:2018zob}.
This signature, however,
is difficult to confirm since most decay modes are prohibited by the kinematical constraint.
A few exceptions are $f_0(1370)\rightarrow \rho\rho$, $f_0(1500)\rightarrow \rho\rho$. These barely satisfy the constraint
through the high tail of their decay widths and, as a result, these partial widths are presumably suppressed substantially.
But some data in PDG show nonnegligible partial widths and the enhancement reported here can give one possible explanation for them.


\section*{Acknowledgments}
This work was supported by the National Research Foundation of Korea(NRF) grant funded by the
Korea government(MSIT) (No. NRF-2018R1A2B6002432, No. NRF-2018R1A5A1025563).


\end{document}